\documentclass[12pt]{article}
\usepackage{t1enc}
\usepackage[utf8]{inputenc}
\usepackage[english]{babel}
\usepackage{graphicx}
\usepackage{amssymb}
\usepackage{amsmath}
\usepackage{amsthm}
\usepackage{braket}

\def\cros{\raise1.9pt\hbox{$\scriptscriptstyle
          >$}\!\raise1.5pt\hbox{$\scriptstyle\triangleleft\,$}}

\def\l{{\lambda}}

\def\w{\wedge}

\theoremstyle{definition}
\theoremstyle{definition}
\theoremstyle{definition}
\theoremstyle{definition}
\newcommand{\noi}{\vspace{0.1in} \noindent}

\title{\bf Two concepts of noncontextuality \\ in quantum mechanics}
\author{\textit{Gábor Hofer-Szabó}\thanks{Research Center for the Humanities, Budapest, email: szabo.gabor@btk.mta.hu}}
\date{}

\begin{document}
\maketitle

\begin{abstract}
There are two different and logically independent concepts of noncontextuality in quantum mechanics. First, an ontological (hidden variable) model for quantum mechanics is called noncontextual if every ontic (hidden) state determines the probability of the outcomes of every measurement independently of what other measurements are simultaneously performed. Second, an ontological model is noncontextual if any two measurements which are represented by the same self-adjoint operator or, equivalently, which have the same probability distribution of outcomes in every quantum state also have the same probability distribution of outcomes in every ontic state. I will call the first concept simultaneous noncontextuality, the second measurement noncontextuality. In the paper I will overview and critically analyze some of the most significant accounts of contextuality in the literature and subsume them under these two categories.
\vspace{0.1in}

\end{abstract}

\section{Introduction}\label{Sec:Int}

On Wayne Myrvold's historical reconstruction,\footnote{I thank Wayne Myrvold for the many valuable information on how the term ``contextuality'' appeared in the foundations of quantum mechanics.} the term ``contextuality'' has been introduced into the foundations of quantum mechanics by Abner Shimony in his contribution to the 1970 Lake Como conference (Shimony 1993, Ch. 7). Shimony first used the somewhat contrived term "contextualistic" which has been shortened to "contextual" in Stuart Freedman's doctoral dissertation, \textit{Experimental Test of Local Hidden-variable Theories} (1972) and has been so used in Beltrametti and Cassinelli's book, \textit{The Logic of Quantum Mechanics} (1981). In his interview for the AIP oral history project\footnote{https://www.aip.org/history-programs/niels-bohr-library/oral-histories/25643.} Shimony ascribes the simplification to Beltrametti and Cassinelli: ``Why didn't I think of that, instead of leaving it to non-English speakers to simplify the words?'' However, Shimony misremembers since  the term  “contextual” already occurred in the Appendix of Clauser and Shimony’s review article (1978, A.2.2). More interestingly, neither of the two papers which are considered as the two seminal works on the subject issue, namely Bell's 1966 paper and Kochen and Specker's 1976 paper uses explicitly the term ``contextuality.'' 

In this paper, however, I am not going to focus on the name of a concept but rather on how the different authors interpreted this concept. The paper consists of two parts. In the first part, I introduce the framework of operational theories and ontological models and accommodate quantum mechanics in this framework (Sect. \ref{Sec:Op}). Then, I discern two concepts of noncontextuality: one based on simultaneous measurability and another based on operational equivalence (Sect. \ref{Sec:NC}). Next, I show how the two concepts connects up to the two different interpretations of the Kochen-Specker argument and how they are related to the ontological, environmental and algebraic contextuality (Sect. \ref{Sec:KS}). In the second half of the paper, I overview some of the most significant accounts of contextuality in the literature and---after discounting contextuality without interpretation (Sect. \ref{Sec:NoC})---I group the accounts and authors into two categories: simultaneous contextuality (Sect. \ref{Sec:SC}) and measurement contextuality (Sect. \ref{Sec:MC}).

In the paper I will argue for the following claims: (i) Simultaneous noncontextuality and measurement noncontextuality are two different and logically independent concepts. (ii) To prove that quantum mechanics does not admit a simultaneous or a measurement noncontextual ontological model requires two different interpretations of the Kochen-Specker theorems. (iii) Formal and the sheaf theoretic approaches are lacking a physical interpretation and hence their contextuality concept is purely mathematical. (iv) Bell 1966 paper excludes simultaneous noncontextual ontological models and Kochen and Specker 1967 paper excludes measurement noncontextual ontological models. (v) Spekkens' and the Bohmian account of noncontextuality is measurement noncontextuality. (vi) Van Fraassen's and Redhead's ontological contextuality is a special case of measurement contextuality; Shimony's algebraic contextuality is a special case of simultaneous contextuality; and Bohr's conceptual contextuality and Redhead's and Shimony's environmental contextuality are uncommitted in the simultaneous-measurement noncontextuality dichotomy.

\section{Operational theories and ontological models}\label{Sec:Op}

An \textit{operational theory} is a theory which specifies the probability of the outcomes of certain measurements performed on a physical system which was previously prepared in certain states. Let $S$ be set of \textit{states} or \textit{preparations} of the system, let $M$ be the set of \textit{measurements} which can be performed on the system, and let $X_m$ be the set of \textit{outcomes} of the measurement $m \in M$. Then, an operational theory is a set of \textit{conditional probabilities} of the outcomes for the various measurements in the various preparations:
\begin{eqnarray}\label{optheory}
\{p(x_m|m\w s) : \, \, x_m \in X_m, \, \, m \in M, \, \, s \in S\}
\end{eqnarray}
with the obvious normalizations. A preparation $s \in S$ is called an \textit{eigenstate} of the measurement $m \in M$ with \textit{eigenvalue} $x_{ms} \in X_m$ if 
\begin{eqnarray} \label{eigenstate}
p(x_{ms}|m\w s) = 1 
\end{eqnarray}

Two measurements $m$ and $m'$ in $M$ are \textit{simultaneously measurable (commeasurable)}, if they can be performed on the same system at the same time. Simultaneous measurability is an empirical question. One can measure the width and the length of a table at the same time but not whether a piece of wood is combustible and whether it floats on water.  Operationally one identifies measurements by sets of (laboratory) instructions. Consequently, two measurements $m$ and $m'$ are simultaneously measurable if and only if there is a measurement which can be identified by the \textit{conjunction} of the sets of instructions characterizing $m$ and $m'$. We call this measurement the \textit{simultaneous measurement} of $m$ and $m'$ and denote it by $m \w m'$ (which is again a measurement in $M$). If $m$ and $m'$ are not simultaneously measurable, we write $m \w m' = \emptyset$. Note that if measurements are identified by instructions and not by their outcome statistics, the simultaneous measurement of $m$ and $m'$ will \textit{not} be any measurement which is---after applying  appropriate functions to the outcomes---"operationally equivalent" (Spekkens, 2005) to $m$ and $m'$.  If it exists, it will be $m \w m'$ defined by the conjunction of the instructions defining $m$ and $m'$.

Denote the set of measurements in $M$ which are commeasurable with $m$ by $M_{\sim m} \subseteq M$. We call an operational theory \textit{non-disturbing} if no conditional probability depends on whether the measurements are performed alone or along with simultaneous measurements:
\begin{eqnarray}\label{nondist}
p(x_m|m\w s)= p(x_m|m\w m'\w s)  \quad \quad x_m \in X_m, \, \, m \in M, \, \, m' \in M_{\sim m}, \, \, s \in S
\end{eqnarray}

Now, quantum mechanics, at least on the minimal interpretation, is a non-disturbing operational theory in a linear algebraic representation. Thus, a physical system is represented by a Hilbert space, each preparation $s \in S$ of the system by a density operator $\boldsymbol{\rho}$, each measurement $m \in M$ by a self-adjoint operator $\textbf{O}_m$, and each outcome $x_m$ of $m$ by the eigenvalue $x_m$ and the associated spectral projection $\textbf{P}_{m}^x$ of $\textbf{O}_m$. The representation is successful if the Born rule holds:  
\begin{eqnarray}\label{born}
\mbox{Tr}(\boldsymbol{\rho}\, \textbf{P}_m^x) = p(x_m|m\w s) \quad \quad x_m \in X_m, \, \, m \in M, \, \, s \in S
\end{eqnarray}
where Tr is the trace function. Note, that the representation of measurements, outcomes and preparations by self-adjoint operators, projections, and density operators, respectively is not unique: all is required from the representation is to satisfy (\ref{born}).

Simultaneous measurements $m$ and $m'$ get represented in quantum mechanics by commuting operators $\textbf{O}_m$ and $\textbf{O}_{m'}$ such that\footnote{For the question as to whether observables represented by noncommuting operators can be simultaneosly measured, see e.g. (Park and Margenau, 1968).}
\begin{eqnarray}\label{born2}
\mbox{Tr}(\boldsymbol{\rho}\, \textbf{P}_m^x  \, \textbf{P}_{m'}^{x'}) = p(x_m \w x_{m'} |m \w m' \w s) \quad \quad x_m \in X_m,  \, \, x_{m'} \in X_{m'}, \, \, s \in S
\end{eqnarray}
Consequently, $m$ and $m'$ will be non-disturbing:
\begin{eqnarray*}
&& p(x_m|m \w m' \w s) = \sum_{x_{m'}} p(x_m \w x_{m'} |m \w m' \w s) = \sum_{x_{m'}} \mbox{Tr}(\boldsymbol{\rho} \, \textbf{P}_m^x \textbf{P}_{m'}^{x'}) =  \nonumber \\
&& \mbox{Tr}(\boldsymbol{\rho} \, \textbf{P}_m^x) = p(x_m|m \w s)  \quad \quad \quad x_m \in X_m, \, \, s \in S
\end{eqnarray*}
That is the quantum mechanical representation of simultaneous measurements by commuting operators implies that quantum mechanics is a non-disturbing (non-signaling) operational theory.

Note that simultaneous measurements as used in this paper are \textit{not} the same as "joint measurements" standardly used in quantum mechanics. By the "joint measurement" of $m$ and $m'$ represented by the commuting operators $\boldsymbol{O}_m$ and $\boldsymbol{O}_{m'}$  in quantum mechanics one usually refers to \textit{any} measurement $m''$ which can be represented by an operator $\boldsymbol{O}_{m''}$ such that there are two Borel functions $f$ and $g$ for which  $\boldsymbol{O}_{m} = f(\boldsymbol{O}_{m''})$ and $\boldsymbol{O}_{m'} = g(\boldsymbol{O}_{m''})$.  This joint measurement $m''$, however, need not be the simultaneous measurement $m \w m'$ defined by the conjunction of the instructions individually characterizing $m$ and $m'$. It can well be any other measurement procedure. On the other hand, the simultaneous measurement $m \w m'$ is a "joint measurement" in the sense that it can be represented by the operator  $\boldsymbol{O}_{m''}$ satisfying the above requirements.

\noi
The role of an \textit{ontological model} (hidden variable model) is to account for the conditional probabilities of an operational theory in terms of underlying realistic entities of the measured system called \textit{ontic states} (hidden variables, elements of reality, beables). Let the set of ontic states be $\Lambda$. An ontological model specifies a \textit{probability distribution} over the ontic states associated with each preparation: 
\begin{eqnarray}\label{ont1}
\{p(\l|s): \, \, \l \in \Lambda, \, \, s \in S\}
\end{eqnarray}
and a set of \textit{response functions} that is a set of conditional probabilities associated with every measurement and every ontic state:
\begin{eqnarray}\label{ont2}
\{p(x_m|m\w \l) : \, \, x_m \in X_m, \, \, m \in M, \, \, \l \in \Lambda \}
\end{eqnarray}
again with the obvious normalizations.

Then, assuming  \textit{no-conspiracy}:
\begin{eqnarray}\label{nocons}
p(\l|s) = p(\l| s \w m)  \quad \quad \, \, m \in M,\, \, s \in S
\end{eqnarray}
(the independence of the probability distributions from the measurements performed on the system) and \textit{$\l$-sufficiency}:
\begin{eqnarray}\label{lsuff}
p(x_m|m\w \l) = p(x_m|m \w \l \w s) \quad \quad  x_m \in X_m, \, \, m \in M,  \, \,  \l \in \Lambda, \, \, s \in S
\end{eqnarray}
(the independence of the response functions from the preparations in which the ontic states are featuring) and using the theorem of total probability one can recover the operational theory from the ontological model in terms of the probability distributions and response functions:
\begin{eqnarray}  \label{recov}
p(x_m|m\w s) &=& \sum_{\l \in \Lambda} p(x_m|m \w \l \w s)  \, p(\l|s \w m) \nonumber \\
&=& \sum_{\l \in \Lambda} p(x_m| m \w \l)  \, p(\l|s)  \quad \quad x_m \in X_m, \, \, m \in M, \, \, s \in S
\end{eqnarray}

Similarly to (\ref{eigenstate}), an ontic state $\l$ is called an \textit{eigenstate} of the measurement $m$ with \textit{eigenvalue} $x_{m\l} \in X_m$ if 
\begin{eqnarray} \label{eigenstate'}
p(x_{m\l}|m\w \l) = 1 
\end{eqnarray} 
An ontological model is called \textit{outcome-deterministic (value-definite)} if every $\l \in \Lambda$ is an eigenstate of every $m \in M$; otherwise it is called \textit{outcome-indeterministic}.

\section{Two concepts of noncontextuality}\label{Sec:NC}

There are two different and logically independent concepts of noncontextuality in quantum mechanics. The first is motivated by the following question: Why is an operational theory non-disturbing? Why simultaneous measurements do not alter the outcome or the probability of an outcome of a measurement in the various \textit{preparations}. A possible answer is to say: because  they don't even alter it in the \textit{ontic states}. This idea is made precise in the concept of noncontextuality:

\noi
\textit{An ontological model is noncontextual if every ontic state determines the probability of the outcomes of every measurement independently of what other measurements are simultaneously performed:}
\begin{eqnarray}\label{NC1}
p(x_m|m \w \l) = p(x_m|m \w m' \w \l) \quad \quad x_m \in X_m, \, \, m \in M, \, \, m' \in M_{\sim m}, \, \, s \in S
\end{eqnarray}
\textit{Otherwise the model is contextual.}

\noi
Thus, noncontextuality is a kind of inference to the best explanation for why an operational theory is non-disturbing: if the ontological model for an operational theory is noncontextual in the sense of (\ref{NC1}), then---assuming no-conspiracy (\ref{nocons}) and $\l$-sufficiency (\ref{lsuff})---one can show that the operational theory is non-disturbing (\ref{nondist}).

But what about disturbing operational theories? 

If two measurements $m$ and $m'$ are disturbing, then a reasonable attitude\footnote{See Bridgman, 1958 and Hofer-Szabó, 2017, Sec. 7.} is to say that the measurements are not yet well-defined. In other words, one should cluster the measurements in a more refined way: instead of taking two measurements: $m$ and $m'$, one should take three measurements: $m_1 \equiv m \w m'$ (measuring both $m$ and $m'$), $m_2 \equiv m \, \w \! \sim \! m'$ (measuring $m$ but not $m'$), and $m_3 \equiv m' \, \w \! \sim \! m$ (measuring $m'$ but not $m$). By this move we can eliminate disturbance since the three new measurements will be logically mutually exclusive. They cannot be simultaneously measured and hence cannot disturb one another.

However, replacing the old measurements by the new ones alters the operational theory. The new fine-grained measurements, being logically mutually exclusive, cannot be simultaneously performed. Hence, noncontextuality cannot be defined since $m_i \w m_j = \emptyset$ on the right hand side of  (\ref{NC1}). In short, in a disturbing operational theory one faces the following dilemma: either we leave the disturbing measurements as they are but then why to require noncontextuality from the ontological model; or we replace the disturbing measurements by mutually exclusive measurements but then noncontextuality will hold vacuously. In both cases noncontextuality becomes meaningless.

Now, the term ``context'' in the above definition is coming from the everyday discourse, where it refers to the \textit{circumstances} in which a certain event or measurement takes place. These circumstances are not constitutive in the definition of the very event or measurement but can significantly influence the occurrence of the event or the result of the measurement. The important aspect of these circumstances is that they are \textit{simultaneously} present with the event or measurement. If I ask you whether you want an ice-cold coke, your answer will greatly depend on whether you are on a sunny beach or on the skating rink \textit{at the time of the question}. A typical context in physics of a measurement is another measurement which is performed simultaneously with the first. In this sense noncontextuality refers to a kind of robustness of a given system to respond in a definite way to a measurement when simultaneous measurements are also performed on the system. 

\noi
There is, however, another meaning of noncontextuality which is completely independent of the above one. This noncontextuality relies not on simultaneous measurability but on the association of measurements with \textit{observables}. The term ``observable'' is missing in the vocabulary of an operational theory since operational theories directly connect measurements to the mathematical representants. But observables are all the more important in standard physical theories. Observables are measurable physical magnitudes which characterize a given physical system. Thus, each observable is associated with a measurement but this association is not necessarily one-to-one. One can measure the same observable in different ways, for example distance by measuring rods and by light signals or temperature by alcohol and gas thermometer. In this case the two measurements have the same outcome or the same probability distribution of outcomes in those preparations in which they both can be performed. In quantum mechanics such measurements are represented by the same self-adjoint operator \textit{via} (\ref{born}). This leads to the second concept of noncontextuality in quantum mechanics which can be motivated by the following question: Why do different measurements have the same probability distribution of outcomes in every \textit{preparation}? Again, a reasonable answer is to say that because they measure the same observable which has a fixed value or a fixed probability distribution of values in every \textit{ontic state} and the faithful measurements simply reveal this value or this distribution. Operationally, this means that the ``operationally equivalent'' (Spekkens, 2005) measurements have the same set of response functions. Thus, the second concept of noncontextuality reads as follows:

\noi
\textit{An ontological model is noncontextual if any two measurements which have the same probability distribution of outcomes in every preparation (and thus can be represented by the same self-adjoint operator) also have the same probability distribution of outcomes in every ontic state. Formally, let $m, m' \in M$ and suppose that}
\begin{eqnarray}\label{NC2a}
p(x_m|m \w s) = p(x_{m'}|m' \w s) \quad \quad x_m \in X_m, \, \,x_{m'}=h(x_m) \in X_{m'}, \, \, s \in S
\end{eqnarray}
where $h:X_m \to X_{m'} $ is a bijection between the outcomes of $m$ and $m'$. Then 
\begin{eqnarray}\label{NC2b}
p(x_m|m \w \l) = p(x_{m'}|m' \w \l) \quad \quad x_m \in X_m, \, \,x_{m'}=h(x_m) \in X_{m'}, \, \, \l \in \Lambda
\end{eqnarray}
\textit{Otherwise the model is contextual.}

\noi
This second noncontextuality is again a kind of inference to the best explanation: (\ref{NC2b})---together with no-conspiracy (\ref{nocons}) and $\l$-sufficiency (\ref{lsuff})---implies (\ref{NC2a}). That is, if an ontological model is noncontextual in the sense of (\ref{NC2a})-(\ref{NC2b}), we obtain a neat explanation for why certain measurements have the same probability distribution of outcomes.

\noi
Let us call the above first concept of noncontextuality (\ref{NC1}) \textit{simultaneous noncontextuality} and the second concept (\ref{NC2a})-(\ref{NC2b}) \textit{measurement noncontextuality} (following Spekkens, 2005). 

Note that measurement noncontextuality and simultaneous noncontextuality are two logically independent concepts. Measurement noncontextuality does not rely on simultaneous measurability, while simultaneous contextuality does. If there are no simultaneous measurements in an operational theory, then each ontological model of the theory will be simultaneous noncontextual since (\ref{NC1}) fulfills vacuously. Still, the model can violate measurement noncontextuality if there are measurements yielding the same probability distribution of outcomes in every preparation, still differing in their response functions.\footnote{For an example see (Hofer-Szabó, 2021a,b).} Conversely, if premise (\ref{NC2a}) is not satisfied in an operational theory, then measurement noncontextuality fulfills vacuously. Still, the ontological model can be simultaneous contextual---but only if the theory is disturbing. In a non-disturbing operational theory for any two simultaneous measurements $m$ and $m'$ (\ref{NC2a}) fulfills for $m$ and $m \w m'$. If the ontological model is measurement noncontextual, then (\ref{NC2b}) also fulfills for $m$ and $m \w m'$. But this is just simultaneous noncontextuality (\ref{NC1}). In short, in a non-disturbing operational theory (like in quantum theory) measurement noncontextuality implies simultaneous noncontextuality.

\section{Noncontextuality in the Kochen-Specker arguments}\label{Sec:KS}

In the previous section measurement noncontextuality and simultaneous noncontextuality were defined at the general level of operational theories and ontological models without any reference to quantum mechanics.\footnote{Modulo the parenthetical note in the definition of measurement noncontextuality which can be dropped.} But quantum mechanics is an operational theory, hence both definitions apply. Thus, one can ask: Is quantum mechanics simultaneous and measurement contextual? 

This question is commonly studied in the Kochen-Specker arguments.\footnote{See (Bell, 1966/2004), (Kochen and Specker, 1967), (Redhead, 1989, Ch. 5) and (Held, 2018).} Kochen-Specker \textit{arguments} have two parts: a Kochen-Specker \textit{theorem} and a physical \textit{interpretation}. The Kochen-Specker \textit{theorems} proceed in the following steps: (i) one starts with a set of self-adjoint operators; (ii) introduces value assignments on the operators sending each operator into one of its eigenvalues; (iii) restricts these value assignments by the so-called functional composition principle such that the values assigned to mutually commuting operators are the eigenvalues of one of the common eigenstates of these operators; and (iv) finally shows that there is no such value assignment. 

The Kochen-Specker arguments are not simply mathematical theorems. They intend to prove that there is no noncontextual outcome-deterministic ontological model underlying quantum mechanics. To this aim, however, the operators in the Kochen-Specker theorems  need to be physically interpreted. An \textit{interpretation} is an association of operators with measurements in an operational theory (\ref{optheory}) and with an underlying ontological model (\ref{ont1})-(\ref{ont2}). 

Now, there are two conceptually different interpretations of the Kochen-Specker theorems. As we will shortly see, on the first interpretation the Kochen-Specker theorems rule out simultaneous noncontextual ontological models, on the second interpretation measurement noncontextual ontological models. The difference between the interpretations in brief is the following: In the first interpretation one directly associates each operator with a measurement in a one-to-one way such that mutually commuting operators represent simultaneous measurements (in the physical sense elucidated in Section \ref{Sec:Op}). In the second interpretation one does not associate each operator with a different measurement but rather each (maximal) set of mutually commuting operators with a measurement. Consequently, operators sitting in more than one such set will be associated with more than one measurement. In short, in  this second interpretation the association of some operators with measurements is of one-to-many type. 

An example might help. Consider the Peres-Mermin-version of the Kochen-Specker theorem. In this theorem one has 9 operators arranged in a 3 $\times$ 3 matrix such that the operators are mutually commuting if and only if they are in the same row or in the same column. Thus, one has 6 triples of mutually commuting operators: the operators in the three rows and in the three columns. Now, the two interpretations differ in the association of the operators with the measurements. In the first interpretation, one associates each operator with a measurement in a one-to-one way such that mutually commuting operators represent simultaneous measurements. Thus, in this operational theory one has altogether 9 different measurements. In the second interpretation, a triple of mutually commuting operators is interpreted not as three different measurements but as three different functions of one and the same measurement. Thus, here one associates not each operator but each triple of mutually commuting operators with a measurement. Thus, in this second operational theory one has altogether 6 different measurements associated with the 6 different triples. The association of operators and measurements, however, is not one-to-one: each operator is associated with two different measurements via the two different commuting triples which it is an element of.

Now, the let's see the two interpretations in a little more detail. In the first interpretation, each ontic state $\l$ is associated with a value assignment $v_\l$ such that the value assigned to an operator ${\bf O}_m$ is an eigenvalue of ${\bf O}_m$ which conforms to the eigenvalue\footnote{See eq. (\ref{eigenstate'}) above.} $x_{m\l}$ in the ontic state $\l$ of the measurement $m$ (uniquely) associated with ${\bf O}_m$:
\begin{eqnarray}\label{value5}
v_\l({\bf O}_m) = x_{m\l} \in X_m
\end{eqnarray}
(\ref{value5}) renders the ontological model outcome-deterministic and also simultaneous noncontextual since $p(x_{m\l}|m \w \l) = 1$ implies that $p(x_{m\l}|m \w m' \w \l) = 1$ for any measurement $m' \in M_{\sim m}$.

If ${\bf O}_m, {\bf O}_{m'}, \dots$ are mutually commuting operators in the Kochen-Specker theorem, then the associated measurements $m, m', \dots$ are simultaneously measurable, that is $m \w m' \w \dots \in M$. The functional composition principle restricts the possible joint outcomes of this simultaneous measurement: only those joint outcomes are allowed which are eigenvalues of one of the common eigenstates of the operators ${\bf O}_m, {\bf O}_{m'}, \dots$.

Now, since there is no value assignment on the Kochen-Specker operators satisfying the functional composition principle, there will be no ontological model satisfying (\ref{value5}). Thus, on the first interpretation, the Kochen-Specker arguments rule out outcome-deterministic \textit{simultaneous noncontextual} ontological models.  

The second interpretation is based on the following mathematical fact: for any set of mutually commuting operators $\{{\bf O}_i\}$ there is a ``fine-grained'' operator ${\bf O}'$ and a set of Borel functions $\{f_i\}$ such that\
\begin{eqnarray}\label{functions}
{\bf O}_i = f_i({\bf O}')
\end{eqnarray}
(The prime indicates that ${\bf O}'$ is not necessarily featuring in the original Kochen-Specker theorem.) Now, the crux of the second interpretation is that one does not directly associate a measurement with every single operator in the set $\{{\bf O}_i\}$; rather one associates a measurement $m'$ only with the fine-grained operator ${\bf O}'$. Consequently, each operator in the set $\{{\bf O}_i\}$ will represent a different function of the very same measurement $m'$. For example, ${\bf O}_i$ will represent $f_i(m')$: ``perform $m'$ and apply the function $f_i$ to the result.''

But then operators sitting in two mutually commuting sets will be ``multiply realized.'' If ${\bf O}_i$ is both in $\{{\bf O}_i\}$ and $\{{\bf O}_j\}$ where $\{{\bf O}_i\}$ and $\{{\bf O}_j\}$ are two different sets of mutually commuting operators, then there will be two different fine-grained operators ${\bf O}'$ and ${\bf O}''$ such that ${\bf O}_i = f_i({\bf O}') = g_i({\bf O}'')$ and hence, if ${\bf O}'$ represents $m'$ and ${\bf O}''$ represents $m''$, ${\bf O}_i$ will represent two different measurements: $f_i(m')$ and $g_i(m'')$.

In this second interpretation, one then assigns values not directly to the Kochen-Specker operators but to the fine-grained operator ${\bf O}', {\bf O}'', \dots$ representing $m, m', \dots$. For the fine-grained operator ${\bf O}'$ the assigned value in the ontic state $\l$ is again the eigenstate $x_{m'\l}$, that is the unique outcome of the measurement $m'$ in the ontic state $\l$:
\begin{eqnarray}\label{value5'}
v_\l({\bf O}') = x_{m'\l} \in X_{m'} 
\end{eqnarray}

Now, (\ref{value5'}) gives a well-defined value assignment on the original Kochen-Specker operators only if each Kochen-Specker operators gets the same value independently of which fine-grained operator it is derived from.  In case of ${\bf O}_i$ this boils down to the requirement that 
\begin{eqnarray}\label{value6'}
f_i(x_{m'\l})=g_i(x_{m''\l})
\end{eqnarray}
But this is exactly measurement noncontextuality (\ref{NC2a})-(\ref{NC2b}): The two measurements $f_i(m')$ and $g_i(m'')$ satisfy the antecedent (\ref{NC2a}) since they are both represented by same operator ${\bf O}_i$ and they also satisfy the consequent (\ref{NC2b}) since from (\ref{value6'}) it follows that:
\begin{eqnarray}\label{FUNC5'}
p(f_i(x_{m'\l})|f_i(m') \w \l) &=& p(g_i(x_{m''\l})|g_i(m'') \w \l)  \quad \quad \l \in \Lambda
\end{eqnarray}

Now, since there is no value assignment satisfying the functional composition principle, there will be no ontological model satisfying (\ref{functions})-(\ref{FUNC5'}). Thus, on this second interpretation, the Kochen-Specker arguments rule out outcome-deterministic \textit{measurement noncontextual} ontological models.

\noi
Some remarks on the two interpretations are in place here: 

\noi
\textit{a}. Note that simultaneous measurability pops up only in the first interpretation. The measurements $\{f_i(m')\}$ represented by $\{{\bf O}_i\}$ in the second interpretation can be called ``simultaneous measurements'' only metaphorically: $m'$ is one single measurement to the result of which one simply applies different functions. Conversely, in the first interpretation there is no mention of measurement noncontextuality since there are no two measurements associated with the same operator.

\noi
\textit{b}. If a Kochen-Specker theorem has a first interpretation, it also has a second interpretation: If each operator ${\bf O}_i$ represents a separate measurement $m_i$ and commuting operators represent simultaneous measurements, then one can associate with each set of mutually commuting operators the simultaneous measurement $m' = \w_i m_i$. Consequently, ${\bf O}_i$ will be associated with all simultaneous measurements which includes $m_i$. 

\noi
\textit{c}. The above interpretations are only the ``pure'' interpretations of the Kochen-Specker theorems. To be sure, one can also provide a ``mixed'' interpretation in which one associates measurements with Kochen-Specker operators in certain commuting sets and with the fine-grained operators in other commuting sets. For example, one can associate the three operators in the first row of the Peres-Mermin square with (three different functions of) one single measurement and uniquely associate the other six operators in the second and third row with six different measurements. Kochen-Specker arguments based on such mixed interpretations, however, neither rule out simultaneous nor measurement noncontextual ontological models since either assumption can be blamed for the no-go result.

\noi
The two interpretations of the Kochen-Specker theorems can also be expressed in terms of observables $\{{\cal O}_i\}$ rather then measurements. The first interpretation is simple: one just uniquely associates each operator with an observable and a measurement. In the second interpretation, however, one needs to choose: either one associates observables with operators or with  measurements in a one-to-one way. If ${\bf O}_i$ represents two measurements $f_i(m')$ and $g_i(m'')$, then one can introduce either \textit{one} observable associated with the operator ${\bf O}_i$ or \textit{two} observables associated with the measurements $m'$ and $m''$. 

In the first case, the operator ${\bf O}_i$ is uniquely associated with an observable $\mathcal{O}_i$ which can be measured in two different ways: either as $f_i(m')$ or as $g_i(m'')$. This is the account of measurement noncontextuality given in Section \ref{Sec:NC}. In the second case, we have two observables $\mathcal{O}'$ and $\mathcal{O}''$ operationally associated with the measurements $m'$ and $m''$, respectively. But then ${\bf O}_i$ will be associated with two observables: $f_i(\mathcal{O}')$ and $g_i(\mathcal{O}'')$. I will refer to this one-to-many association of operators and observables as \textit{general ontological contextuality}. In the literature,\footnote{See Subsection \ref{Subsec:vFC}.} ontological contextuality is used in a narrower sense, namely when the association of operators and observables is one-to-one for maximal (non-degenerate) operators but one-to-many for nonmaximal (degenerate) operators. I will call this latter type of contextuality \textit{special ontological contextuality}. All in all, ontological contextuality is just the sum of two facts: the existence of operationally equivalent measurements \textit{plus} the operational association of observables with measurements.

The two cases lead to two different accounts of measurement contextuality \textit{in terms of observables}. Define the \textit{(measured) value of an observable} $\mathcal{O}$ in the ontic state $\l$ as the eigenvalue $x_{m\l}$ of the measurement $m$ associated with the observable $\mathcal{O}$ in the ontic state $\l$. Denote this value by $[\mathcal{O}]_\l$. Now, in the first case one has only one observable $\mathcal{O}_i$; its value, however, will depend on which measurement it is measured with. In the second case one has two observables $\mathcal{O}'$ and $\mathcal{O}''$ with values $x_{m'\l}$ and $x_{m''\l}$, respectively. Thus, in the first case the violation of measurement noncontextuality (\ref{value6'}) means that 
\begin{eqnarray}\label{envcont}
[\mathcal{O}_i]_\l(f_i(m')) \neq [\mathcal{O}_i]_\l(g_i(m''))
\end{eqnarray}
that is the value of $\mathcal{O}_i$ in a fixed ontic state depends on whether the observable is measured via $f_i(m')$ or $g_i(m'')$. This dependence is a special case of \textit{environmental contextuality}. Environmental contextuality is the dependence of the outcome of a measurement or the value of an observable on the environment of the system. Depending on how the term ``environment'' is characterized, environmental contextuality can mean many different things. In the literature ``environment'' typically means: (i) the measurement itself with which the observable is measured, (ii) the macrostate of the measuring apparatus, (iii) the presence or absence of other simultaneously performed measurements, etc. (\ref{envcont}) is environmental contextuality in the sense of (i).  

In the second case, the meaning of measurement contextuality is much simpler: due to ontological contextuality $\mathcal{O}'$ and $\mathcal{O}''$ are two different observables, therefore it is not surprising that in a given ontic state their value differ:
\begin{eqnarray}\label{ontcont}
f_i([\mathcal{O}']_\l) \neq g_i([\mathcal{O}'']_\l)
\end{eqnarray}

To sum up, the two different associations of observables with operators and measurements lead to two different accounts of measurement contextuality in terms of observables, the one based on environmental contextuality in the sense of (i), the other on ontological contextuality.

Finally, note that simultaneous contextuality is also a special case of environmental contextuality but in the sense of (iii). Following Shimony (2008) I will call environmental contextuality in the sense of (iii) \textit{algebraic contextuality}.\footnote{See Subsection \ref{Subsec:vFC}.}  Also note that simultaneous contextuality is not related in any sense to ontological contextuality.

\section{Contextuality without interpretation}\label{Sec:NoC}

In this second part of the paper I will overview some of the most significant accounts of contextuality in the literature and cast them into two groups: simultaneous contextuality (Sections \ref{Sec:SC}) and measurement contextuality (Sections \ref{Sec:MC}). In categorizing the authors I will be careful not to rely too much on how they use the term ``simultaneous measurements'' since in the physicists' parlance ``simultaneous measurements'' and ``commuting operators'' are often used synonymously. Rather, I will identify the position of the authors on the basis of their interpretation of the Kochen-Specker theorems. Namely, if in interpreting a Kochen-Specker theorem they uniquely associate  operators with measurements (observables) and declare explicitly or think implicitly that commuting operators represent simultaneous measurement, then they presumably fall into the first group. If they associate some sets of mutually commuting operators with one single measurement (observable) and consequently they associate some operators with more than one measurement, then they supposedly fall into the second group. 

Let me make it clear, however, that the upcoming discussion is not meant as a comprehensive historical analysis of contextuality in quantum theory. Rather it is an attempt to ``course-grain'' the most prominent authors into two groups around the two different concepts contextuality. 

Before, however, I need to separate a third type of contextuality from the above two which I will still call, for lack of a better word, contextuality: contextuality without interpretation. Note that both simultaneous and measurement noncontextuality were based on a physical interpretation of quantum mechanics that is an association of the operators (operationally defined) measurements. In the absence of such an association contextuality remains uninterpreted, that is a pure mathematical concept. The upcoming two approaches provide different examples to such a formal understanding of contextuality.

\subsection{Formal contextuality}\label{Subsec:FC}

In the mathematically oriented approach to the foundations of quantum mechanics\footnote{Especially in quantum information, see e.g. Klyachko et al. 2008; Bub and Stairs, 2009; Cabello, 2013.} contextuality is often treated at the purely formal level. In this approach contextuality is simply identified with the construction of a Kochen-Specker \textit{theorem} that is the impossibility to provide a value assignment on a tricky set of self-adjoint operators satisfying the functional composition principle.

Needless to say, that if the operators featuring in these theorems (along with commutativity, value assignment, functional composition principle, etc.) are not interpreted physically, then contextuality remains a purely mathematical concept without any reference to the outer world. Sometimes this lack of interpretation is concealed by an inadequate terminology which calls operators  ``observables,'' projections ``events,'' commutativity ``simultaneous measurability,'' and a complete set of orthogonal projections a ``context.''  These Janus-faced expressions might contribute to blur the boundaries between mathematics and physics and convey the impression as if the formalism automatically brought with it a physical interpretation. However, if the goal is to prove that nature is contextual, then the physical interpretation of these terms needs to be given explicitly. Note, however, that when I stick to a physical interpretation I don't expect a full-fledged interpretation of quantum mechanics but simply a \textit{minimal} one where operators are associated with operationally defined measurements.

\subsection{Sheaf theory} 

There is an elegant treatment of contextuality, the \textit{sheaf theoretic} approach.\footnote{See Isham and Butterfield, 1998; Butterfield and Isham, 1999; Abramsky et al. 2011, 2017.} The basic ingredients of the approach are: the set of measurements $M$ and the set of outcomes $X$; the set of contexts $\{C_k\} = \big\{\!\{m_{i_k}\}\!\big\}$ that is the set of maximal subsets of \textit{compatible} measurements; and for each context $C_k$ a probability distribution $p_k$ on the set of functions: $C_k \to X$. The probability distributions are consistent in the sense that for any two contexts $C_k$ and $C_l$, the distributions $p_k$ and $p_l$ marginalize to the same distribution on $C_k \w C_l$. These probability distributions are said to satisfy the \textit{sheaf condition} if there is  a probability distribution $d$ on $M$, called global section, which marginalizes to the distribution $d_k$ on each contexts $C_k$. One of the theory's goal is to set up conditions on which a set of probability distributions on contexts can have a global section.

Now, \textit{if} measurements, outcomes, etc. in the sheaf theory represent physical measurement, their outcomes, etc. and compatibility means simultaneous measurability, then the above formalism boils down to an operational theory (\ref{optheory}) in a fixed preparation and the sheaf condition provides a necessary and sufficient condition for the operational theory to have a simultaneous noncontextual outcome-deterministic (or more generally a factorizing\footnote{Noncontextual non-factorizing ontological models are not excluded by the sheaf-theoretic arguments (and neither by the Kochen-Specker arguments). See for example Spekkens 2005, Sec. 8.B.}) ontological model. 

However, sheaf-theory, as a mathematical theory, leaves the compatibility relation uninterpreted. Moreover, compatibility is intentionally devised to be so general as to cover a wide range of mathematical and physical interpretations: it can represent commutativity in quantum theory, consistency in database theory, simultaneous measurement in operational theories, etc. As Abramsky et al. put it:
\begin{quote}
It should be noted that measurement covers provide a very general way of representing compatibility relationships. Of course, a physical interpretation in particular circumstances will give rise to specific structures of this kind. (Abramsky et al. 2011, p. 8)
\end{quote}
In this respect, it is very instructive to see how Christopher Isham and Jeremy Butterfield, the authors of one the programmatic papers on the sheaf (topos) theoretic approach to quantum mechanics explicitly opt for a formal treatment relying on operators rather than a physical approach relying on physical observables: 
\begin{quote}
In a quantum theory, a physical quantity $A$ is represented by a self-adjoint operator $\bf{A}$ on the Hilbert space of the system, and the first thing one has to decide is whether to regard a valuation as a function of the physical quantities themselves, or on the operators that represent them. From a mathematical perspective, the latter strategy is preferable, and we shall therefore define a (global) valuation to be a real-valued function $V$ on the set of all bounded, self-adjoint operators. (Isham and Butterfield, 1998, p. 2670)
\end{quote}
And they go on to study value assignments to self-adjoint operators. This move is very similar to the Kochen-Specker \textit{theorems} where the point is to provide a coloring no-go result on a set of \textit{operators}. But as we saw before, neither Kochen-Specker theorems nor  sheaf theory \textit{per se} can prove quantum contextuality until the mathematical terms featuring in the arguments are not interpreted physically.

But now let's turn to the two contextuality concepts \textit{with} physical interpretation.

\section{Simultaneous contextuality}\label{Sec:SC}

\subsection{Bell and Shimony}\label{Subsec:BSC}

In his seminal paper reviewing the hidden variable program Bell (1966) provided a no-go theorem against outcome-deterministic noncontextual ontological models for quantum mechanics (without uttering the word ``contextual''). Bell's no-go result is based on three one-dimensional projections, the first two of which are orthogonal to the third but not to one another (therefore the argument works only in at least three dimensions). Bell assigns $0$ or $1$ values to these projections; applies the sum rule\footnote{A special case of the functional composition principle which requires that the values assigned to a complete set of orthogonal projections should sum up to $1$.} to these projections; and derives a contradiction. In interpreting the contradiction, Bell is wondering on how the sum rule, ``an assumption in which only commuting operators were explicitly mentioned,'' can lead to a no-go result making also use of noncommuting operators. He concludes:
\begin{quote}
The danger in fact was not in the explicit but in the implicit assumptions. It was tacitly assumed that measurement of an observable must yield the same value independently of what other measurements may be made simultaneously. (Bell, 1966/2004, p. 9)
\end{quote}
It becomes clear from Bell's wording that he uses the term ``simultaneous measurements'' as a synonym for ``commuting operators.''\footnote{At the beginning of his paper Bell is more explicit on this: ``Observables with commuting operators can be measured simultaneously'' (Bell, 1966/2004, p. 2).} The observable represented by the third projection can be simultaneously measured with the observable represented by the first or second orthogonal projection but not with both. 
\begin{quote}
These different possibilities require different experimental arrangements; there is no \textit{a priori} reason to believe that the results for [the third projection] should be the same. (Bell, 1966/2004, p. 9)
\end{quote}
That is measuring the third observable simultaneously with the first or with the second observable (or just alone) requires two (three) physically different measurements settings and one can avoid the contradiction by assigning different values to the third observable depending on which measurement it is simultaneously measured with. 

In other words, Bell interprets his own no-go result along the lines of the one operator-one measurement interpretation: he associates each of the three projections with a different observable (and a corresponding measurement) such that commuting operators represent simultaneous measurements. On this interpretation, his no-go results rule out \textit{simultaneous} noncontextual (outcome-deterministic) ontological models for quantum theory. This reading of Bell's understanding of noncontextuality is all the more consistent with his locality concept as the robustness of the system' response to a measurement against remote simultaneous measurements.

Following Bell, many authors working in the foundations of quantum mechanics interpret noncontextuality as simultaneous noncontextuality. Here I mention only Abner Shimony's characterization of a noncontextual (outcome-deterministic) hidden variable model:\footnote{See also Shimony, 1984, p. 36.}
\begin{quote}
A noncontextual hidden variables model postulated that an isolated physical system is characterized by a complete state $\l$, which is the compendium of the real properties of the system at a definite time\dots When $\l$ is given, then the result of measuring any property $A$ of the system at the given time by an ideal measuring apparatus\dots is a function $A(\l)$. The outcome of the measurement is assumed to be independent of other properties $B$, $C$, \dots that may be measured simultaneously with $A$, and indeed such an independence may be tacitly assumed to be intrinsic to the ideal character of the measurement process. (Shimony, 2009, p. 289)
\end{quote}

\subsection{Bohr's conceptual, Redhead's environmental and Shimony's algebraic contextuality}\label{Subsec:BRS}

Bell explicates simultaneous contextuality as follows:
\begin{quote}
The result of an observation may reasonably depend not only on the state of the system (including hidden variables) but also on the complete disposition of the apparatus. (Bell, 1966/2004, p. 9)
\end{quote}
Thus, Bell interprets simultaneous contextuality as a special case of environmental contextuality, that is the impact of the measurement apparatus on the system. This impact can be understood in two different ways: either as a causal influence or a kind of conceptual or definitional relation. A prominent representative of this latter type of influence is Niels Bohr to whom Bell is referring at the end of his lastly quoted passage.\footnote{\textit{Cf}. ``Bell, by a judo-like maneuverer, cited Bohr in order to vindicate a family of hidden variables theories in which the values of observables depend not only upon the state of the system but also upon the context.'' (Shimony, 1984, p. 41)} Here Bohr draws attention to
\begin{quote}
\dots the impossibility of any sharp distinction between the behavior of atomic objects and the interaction with the measuring instruments which serve to define the conditions under which the phenomena appear. (Bohr, 1949)
\end{quote}
In another quote Bohr is more explicit on this point:
\begin{quote}
\dots the procedure of measurement has an essential influence on the conditions on which the very definition of the physical quantities in question rests. (Bohr, 1935)
\end{quote} 

Most authors, however, opt for a causal reading of the influence of the measuring apparatus on the system. Heywood and Redhead (1983) and Redhead (1989), for example, write:
\begin{quote}
[T]here is some nonquantum interaction between the system of interest and its surroundings  which occurs before the act of measurement and alters the values of magnitudes of the system. (Heywood and Redhead, 1983 p. 488; Redhead, 1989 p. 140)\footnote{The timing of the interaction is not important: ``presumably there is no reason why they should only occur immediately prior to a measurement.'' (Heywood and Redhead, 1983 p. 488; Redhead, 1989 p. 140)} 
\end{quote}
Shimony (1984) characterizes environmental contextuality more generally:
\begin{quote}
In practice, of course, the environment is very imprecisely known and characterized. Consequently there is motivation to let $C$ [the context] be a 'coarse-grained state' of the environment, such as is specified by a macroscopic description of the apparatus.'' (Shimony, 1984 p. 29)
\end{quote}

At the end of Section \ref{Sec:KS} I discerned three different meanings of environmental contextuality. Clearly, Redhead and Shimony (and maybe Bohr) employ the term ``environmental contextuality'' in the sense of (ii) which is not related to either simultaneous or measurement contextuality. I called environmental contextuality in the sense of (iii) algebraic contextuality. This name goes back to Shimony's (2008)\footnote{See also (Shimony, 1984, p. 29) and (Gudder 1970).} definition of an algebraic context:
\begin{quote}
There actually are two quite different versions of contextual hidden-variables theories, depending upon the character of the context: an “algebraic context” is one which specifies the quantities (or the operators representing them) which are measured jointly with the quantity (or operator) of primary interest, whereas an “environmental context” is a specification of the physical characteristics of the measuring apparatus whereby it simultaneously measures several distinct co-measurable quantities.
\end{quote}
In this quote Shimony contrasts algebraic and environmental contextuality, whereas I find it more proper to take the former to be a special case of latter since a simultaneous measurement is also special environment. But be that as it may, Shimony's algebraic contextuality is a special case of simultaneous contextuality (\ref{NC1}).

\section{Measurement contextuality}\label{Sec:MC}

\subsection{Kochen and Specker} 

In their pioneering paper, Kochen and Specker (1967) showed that there is no value assignment on a set of 117 projections in a 3-dimensional Hilbert space which respects the sum rule on every triplet of orthogonal projections. Each triplet of projections are squared spin-1 operators associated with three orthogonal spatial directions. What is the physical interpretation of these projections?

The projections are commuting but Kochen and Specker are careful in distinguishing the commutativity of the operators and the simultaneous measurability (``commeasurability'') of the represented measurements. They write:
\begin{quote}
Of course, we have seen that these operators commute and it is a generally accepted assumption of quantum mechanics that commuting operators correspond to commeasurable observables. A rationale for this assumption \dots is that if ${\bf A}_i, i \in I$ is a set of mutually pairwise commuting self-adjoint operators, then there exists a self-adjoint operator ${\bf B}$ and Borel functions $f_i, i\in I$ such that ${\bf A}_i = f_i(\bf{B})$. However, this justification hinges on the existence of a physical observable which corresponds to the operator. (Kochen and Specker, 1967, 72)
\end{quote}
What Kochen and Specker state here is that from the mathematical fact that for a set of mutually commuting operators $\{{\bf A}_i\}$ one can find an operator ${\bf B}$ such that ${\bf A}_i = f_i(\bf{B})$, it does not follow that there also is a \textit{physical measurement} realizing ${\bf B}$. Therefore, they construct for every triplet of projections a self-adjoint operator, the spin-Hamiltonian, and three functions. The spin-Hamiltonian represents a physical observable: ``the change in the energy of the lowest orbital state of orthohelium resulting from the application of a small electric field with rhombic symmetry'' (Kochen and Specker, 1967, 73). They also associate a measurement to this observable: the measurement of energy of the orthohelium. 

What is important is that Kochen and Specker do \textit{not} interpret the triplets of spin operators as \textit{three different} spin measurements on a spin-1 particle. These measurements would not be simultaneous measurable. Instead they associate with each triplet one single energy measurement performed on a system of orthohelium perturbed by a rhombic electric field. When they speak about ``simultaneous measurement'' of the three spin \textit{operators} (p. 73), they mean exactly this energy measurement. Therefore, I think, they use the term ``simultaneous measurement'' only metaphorically,\footnote{See Remark \textit{a} in Section \ref{Sec:KS}.} namely performing one single measurement and, based on the outcome, assigning values to the projections.

But if each triplet is associated with a different energy measurement, then projection popping up in two different triplets will be associated with two different measurements. Consequently, on the authors own interpretation the Kochen-Specker theorem rules out \textit{measurement} noncontextual (outcome-deterministic) ontological models for quantum mechanics.

\subsection{Van Fraassen's and Redhead's ontological contextuality}\label{Subsec:vFC}

The concept (but not the name) of ontological contextuality goes back to Bas van Fraassen. In his 1973 paper he writes:
\begin{quote}
[M]aximal observables can be identified with the operators that represent them. All other observables are functions of maximal observables, and are identified jointly by a maximal observable and a function. (Van Fraassen, 1973, p. 107)
\end{quote}
At the end of Section \ref{Sec:KS} I called this concept  \textit{special ontological contextuality} where the non-unique association of operators and observables is confined only to non-maximal operators. The general concept, which I called \textit{general ontological contextuality}, has been formulated in van Fraassen's 1979 paper:
\begin{quote}
[t]wo observables [$a$ and $b$] are statistically equivalent if they have the same probability distribution\dots In that case they are represented in physics by the same Hermitean operator\dots But that does not mean that $a=b$. (Van Fraassen, 1979, p. 158)
\end{quote}
In other words, two observables can be represented by the same self-adjoint operator without being the same. But then, one is not compelled to assign the same value to the observables or the same response function to the associated measurements in the Kochen-Specker argument. This a clear case of measurement contextuality. 

Michael Redhead picks up van Fraassen's view in analyzing the Kochen-Specker argument:
\begin{quote}
[O]bservables corresponding to maximal operators are ontologically prior to those which correspond to nonmaximal operators. Knowing to which self-adjoint operator it corresponds is not sufficient to identify unambiguously a nonmaximal observable: we must know also to which maximal observable its values are related. It requires in this sense a context; and this splitting of nonmaximal observables, so that each nonmaximal operator now corresponds to many distinct observables, we shall refer to as \textit{Ontological Contextuality}. Redhead (1989, p. 135)
\end{quote}
Note, that since to block the Kochen-Specker arguments, it is enough to confine ontological contextuality to nonmaximal operators, therefore Redhead uses the special version of ontological contextuality.

\subsection{Spekkens' measurement contextuality} 

In the operational approach to quantum mechanics initiated by Robert Spekkens (2005) noncontextuality is defined as follows:
\begin{quote}
A noncontextual ontological model of an operational theory is one wherein if two experimental procedures are operationally equivalent, then they have equivalent representations in the ontological model. (Spekkens, 2005, p. 1) 
\end{quote}
If the ``experimental procedure'' is a measurement, then Spekkens' noncontextuality boils down to measurement noncontextuality (\ref{NC2a})-(\ref{NC2b}).\footnote{Spekkens applies the concept of noncontextuality also to preparations and transformations.}  Many authors working in the operational approach follow this measurement account of noncontextuality.\footnote{See e.g. Hermens, 2011; Liang et al., 2011; Leifer, 2014; Krishna et al., 2017.} There are also experiments devised to test noncontextuality in this sense.\footnote{Mazurek, 2016.}

\subsection{Bohmian mechanics} 

Bohmian mechanics is an outcome-deterministic measurement contextual ontological model for quantum mechanics. The classical example for the violation of measurement noncontextuality (\ref{NC2a})-(\ref{NC2b}) is due to Albert (1992, p. 150).\footnote{See also Hemmick and Shakur, 2012, Ch. 2.5 and 2.6; Dürr et al., 2013, Ch. 3.8.3.} Consider two scenarios of spin measurement  performed on an electron which differ only in the polarity of the Stern-Gerlach magnets. If one identifies the spin outcomes associated with the upper path in the first scenario with the spin outcomes associated with the lower path in the second scenario, then the two spin measurements will have the same distribution of outcomes in every quantum state and hence can be represented by the same spin operator in quantum mechanics. Still, they provide opposite spin outcomes for a fixed ontic state in Bohmian mechanics. Namely, in Bohmian mechanics the ontic states are just the possible initial positions of the electron together with the wave function of the electron. 
If the measurement situation has a certain reflection symmetry with respect to a plane between the Stern-Gerlach magnets, then the Bohmian equations of motion ensure that the electron cannot cross the symmetry plane of the measurement. This means that electrons starting from above the plane will remain above the plane in \textit{both} measurement situations. But then the two measurements will give opposite spin outcomes for one and the same ontic state. Thus, measurement noncontextuality (\ref{NC2a})-(\ref{NC2b}) is violated.

There is, however, another meaning of contextuality often used with respect to Bohmian mechanics. It is said that the spin (and all other observables except position) is \textit{contextual}---meaning that it depends not only on the state of the system but also on the state of the measurement apparatus. In other words, spin is not an inherent property of the system rather a kind of coarse-graining over the final positions of the electron after the Stern-Gerlach measurement. This is environmental contextuality in the sense of (ii).\footnote{See the end of Section \ref{Sec:KS}.}

Some Bohmians object against calling this latter fact contextuality. As Daumer et al. put it:
\begin{quote}
We thus believe that contextuality reflects little more than the rather obvious observation that the result of an experiment should depend upon how it is performed! (Daumer et al., 1996, p. 389)
\end{quote}
I agree with these authors: measurement outcomes need not depend only on the operator associated with a measurement but can also depend on the very measurement. In this sense there is nothing extraordinary in that Bohmian mechanics (or any other physical theory) is contextual. However, Bohmian mechanics also provides a striking example for an ontological model in which the ontic states respond differently to different measurements represented by the same operator. That is Bohmian mechanics provides an example for a measurement contextual ontological model. The example is striking since measurement noncontextuality, as said in Section \ref{Sec:NC}, is a kind of inference to the best explanation for why different measurements can have the same distribution of outcomes in every preparation. It is not an a priori truth but, no doubt, a reasonable requirement on ontological models. And Bohmian mechanics shows how an ontological model can still live without it.

\section{Conclusions}

Simultaneous noncontextuality and measurement noncontextuality are two different and logically independent concepts. To prove that quantum mechanics is contextual in the one sense or in the other requires different strategies that is different interpretations of the operators in a Kochen-Specker theorem. To prove that quantum mechanics is simultaneous contextual, one needs to associate the Kochen-Specker operators with physical measurements in a one-to-one way. To prove that quantum mechanics is measurement contextual, such a one-to-one association is not needed: operators sitting in more than one commuting subsets can be associated with more than one measurements. Different authors in the literature using the same Kochen-Specker theorems provide different Kochen-Specker arguments depending on whether they use the one operator-one measurement interpretation or the one operator-many measurements interpretation. Consequently they prove quantum contextuality in two different senses. 

In the paper I grouped some of the most prominent accounts on contextuality into two categories: simultaneous contextuality and measurement contextuality. I argued that the formal and the sheaf theoretic approaches are lacking a physical interpretation and hence their contextuality concept is purely mathematical. I also argued that the two seminal papers, namely Bell (1966) and Kochen and Specker (1967) provide two different interpretations of a no-go theorem and hence the former excludes simultaneous noncontextual, the latter measurement noncontextual ontological models. Next, I argued that Spekkens' and the Bohmian account of noncontextuality is measurement noncontextuality (which in a non-disturbing theory like quantum theory includes simultaneous noncontextuality). Finally, I argued that Van Fraassen's and Redhead's ontological contextuality is a special case of measurement contextuality; that Shimony's algebraic contextuality is a special case of simultaneous contextuality; and that Bohr's conceptual contextuality and Redhead's and Shimony's environmental contextuality are unrelated to either simultaneous or measurement contextuality.
\vspace{0.2in}

\noindent
{\bf Acknowledgements.} This work has been supported by the Alexander von Humboldt Foundation, the National Research, Development and Innovation Office (K-115593), and a Senior Research Scholarship of the Institute of Advanced Studies Koszeg. I thank Márton Gömöri for his careful reading of the manuscript.

\section*{References}
\footnotesize

\begin{list} 
{ }{\setlength{\itemindent}{-15pt}
\setlength{\leftmargin}{15pt}}

\item S. Abramsky and A. Brandenburger, ''The sheaf-theoretic structure of non-locality and contextuality,'' \textit{New Journal of Physics}, \textbf{13}, 113036 (2011).

\item S. Abramsky, R. S. Barbosa, K. Kishida, R. Lal, S. Mansfield, ''Contextuality, Cohomology and Paradox,'' URL = https://arxiv.org/abs/1502.03097 (2017).

\item D. Z. Albert, \textit{Quantum Mechanics and Experience}, (Cambridge Massachusetts: Harvard University Press, 1992).

\item J. S. Bell, "On the problem of hidden variables in quantum mechanics", \textit{Reviews of Modern Physics}, \textbf{38}, 447-452 (1966) reprinted in J. S. Bell, \textit{Speakable and Unspeakable in Quantum Mechanics}, (Cambridge: Cambridge University Press, 2004).

\item P. W. Bridgman, \textit{The Logic of Modern Physics}, (New York: The Macmillan Company, 1958/1927).

\item J. Bub and A. Stairs, ''Contextuality and Nonlocality in ‘No Signaling’ Theories,'' \textit{Foundations of Physics}, \textbf{39 (7)},   690-711 (2009).

\item A. Cabello, ''Twin inequality for fully contextual quantum correlations,'' \textit{Physical Review A}, \textbf{87}, 010104(R) (2013).

\item J. Butterfield and C. J. Isham, ''Topos perspective on the Kochen-Specker Theorem: II. Conceptual aspects and classical analogues,'' \textit{International Journal of Theoretical Physics}, \textbf{38}, 827–859 (1999).

\item  J. F. Clauser and A. Shimony, ``Bell's theorem. Experimental tests and implications,'' \textit{Reports on Progress in Physics}, \textbf{41 (12)}, 1881-1927 (1978).

\item M. Daumer, D. Dürr, S. Goldstein, and N. Zanghì, ''Naive Realism about Operators,'' \textit{Erkenntnis}, \textbf{45},  379-397 (1997).

\item D. Dürr, S. Goldstein, and N. Zanghì, \textit{Quantum Physics without Quantum Philosophy}, (Dordrecht: Springer Verlag, 2013).

\item D. Greenberger, M. Horne, and A. Zeilinger, ''Going beyond Bell’s theorem,'' in M. Kafatos (ed.), \textit{Bell’s Theorem, Quantum Theory, and Conceptions of the Universe} (Kluwer Academic, Dordrecht, 1989), 69–72.

\item S. Gudder, ''On hidden-variable theories,'' \textit{Journal of Mathematical Physics}, \textbf{11}, 431-6 (1970).


\item C. Held, ''The Kochen-Specker Theorem,'' \textit{Stanford Encyclopedia of Philosophy},

URL = https: //plato.stanford.edu/entries/kochen-specker/ (2018).

\item D. L. Hemmick, and A. M. Shakur, \textit{Bell's Theorem and Quantum Realism -- Reassessment in Light of the Schrödinger Paradox}, (Dordrecht: SpringerBriefs in Physics, 2012).

\item R. Hermens, ''The problem of contextuality and the impossibility of experimental metaphysics thereof,'' \textit{Studies in History and Philosophy of Modern Physics}, \textbf{42 (4)}, 214-225 (2011).

\item P. Heywood and M. Redhead, ''Nonlocality and the Kochen-Specker paradox,'' \textit{Foundations of Physics}, \textbf{13/5}, 481-499 (1983).

\item G. Hofer-Szabó, ''How man and nature shake hands: the role of no-conspiracy in physical theories,'' \textit{Studies in History and Philosophy of Modern Physics}, \textbf{57}, 89-97 (2017).

\item G. Hofer-Szabó, ''Commutativity, comeasurability, and contextuality in the Kochen-Specker arguments,'' \textit{Philosophy of Science}, \textbf{ 88}, 483-510. (2021a).

\item G. Hofer-Szabó, ''Three noncontextual hidden variable models for the Peres-Mermin square,'' \textit{European Journal for the Philosophy of Science}, \textbf{ 11}, 30. (2021b).

\item A. Klyachko, M. A. Can, S. Binicioğlu, and A. S. Shumovsky, ``A simple test for hidden variables in spin-1 system,'' \textit{Phys. Rev. Lett}. \textbf{101}, 020403 (2008).

\item S. Kochen and E.P. Specker, "The problem of hidden variables in quantum mechanics", \textit{Journal of Mathematics and Mechanics}, \textbf{17}, 59–87 (1967).

\item A. Krishna, R. W. Spekkens, and E. Wolfe, ''Deriving robust noncontextuality inequalities from algebraic proofs of the Kochen-Specker theorem: the Peres-Mermin square,'' \textit{New Journal of Physics}, \textbf{19}, (2017).

\item C. J. Isham and J. Butterfield, ''Topos perspective on the Kochen-Specker Theorem: I. Quantum states as generalized valuations,'' \textit{International Journal of Theoretical Physics}, \textbf{37}, 2669–2733 (1998).

\item M. Leifer, ''Is the quantum state real? An extended review of $\psi$-ontology theorems,'' \textit{Quanta.} \textbf{3(1)} 67-155 (2014). 

\item Y. Liang, R. W.Spekkens, H. M. Wisemand, ''Specker’s parable of the overprotective seer: A road to contextuality, nonlocality and complementarity,'' \textit{Physics Reports}, \textbf{506 (1-2)}, 1-39 (2011).


\item J. L. Park, and H. Margenau, ''Simultaneous Measurability in Quantum Theory,'' \textit{International Journal of Theoretical Physics}, \textbf{1}, 211 (1968).

\item M. Redhead, \textit{Incompleteness, Nonlocality, and Realism}, (Oxford: Oxford University Press, 1989).

\item A. Shimony, ''Contextual hidden variables theories and Bell's inequalities,'' \textit{The British Journal for the Philosophy of Science}, \textbf{35}, 25-45 (1984).

\item A. Shimony, ``Bell's theorem,'' \textit{Stanford Encyclopedia of Philosophy}

URL = https://stanford.library.sydney.edu.au/archives/fall2008/entries/bell-theorem/


\item A. Shimony, \textit{Search for a Naturalistic World View}, Book 2, (Cambridge: Cambridge University Press, 1993).

\item A. Shimony, ''Hidden-variables models of quantum mechanics (noncontextual and contextual),'' in: D Greenberger, K. Hentschel, and F. Weinert (eds.) \textit{Compendium of Quantum Physics}, (Berlin: Springer-Verlag, 2009), 287-291.

\item R. W. Spekkens, ''Contextuality for preparations, transformations, and unsharp measurements,'' \textit{Physical Review A}, \textbf{71},   052108 (2005).

\item B. C. Van Fraassen, ''Semantic Analysis of Quantum Logic,'' in C.A. Hooker (ed.), \textit{Contemporary Research in the Foundations and Philosophy of Quantum Theory}, (Dordrecht: Reidel, 1973).

\item B. C. Van Fraassen, ''Hidden variables and the modal interpretation of quantum theory,'' \textit{Synthese}, \textbf{42}, 155-65 (1979).


\end{list}
\end{document}